%Paper: 9111012
%From: minahan@gomez.phys.virginia.edu
%Date: Tue, 05 Nov 1991 10:33:03 EST

%%There is a postscript file attached to the end of this file which contains
%%four figures.
%% Peel off this file at the line marked "Cut postscript file here"

\input phyzzx

\Pubnum={UVA-HET-91-07}
\date={October 1991}
%\pubtype={}
\titlepage
\title{Flows and Solitary Waves in Unitary Matrix Models\break
with Logarithmic Potentials.\footnote*
{Supported in part by the United States
Department of Energy under grant DE-AS05-85ER-40518.}}
\bigskip
\author {Joseph~A.~Minahan\footnote\dag
{Electronic mail:  MINAHAN@gomez.phys.virginia.edu}}
\address{Deparment of Physics,  Jesse Beams Laboratory,\break
University of Virginia, Charlottesville, VA 22901 USA}
\bigskip
\abstract{We investigate unitary one-matrix models coupled to bosonic
quarks.  We derive a flow equation for the square-root of the
specific heat as a function of the renormalized quark mass.
We show numerically that the flows have a finite number of
solitary waves, and we postulate that their number equals
the number of quark flavors.
We also study the nonperturbative behavior of this theory and show that as the
number of flavors diverges, the flow does not reach two-dimensional gravity.}
\endpage

\def\NP{{\it Nucl. Phys.\ }}
\def\PL{{\it Phys. Lett.\ }}
\def\PRD{{\it Phys. Rev. D\ }}
\def\PRL{{\it Phys. Rev. Lett.\ }}
\def\CMP{{\it Comm. Math. Phys.\ }}

\def\g{\gamma}
\def\gg{\g+1/\g}
\def\Rn{R_n}
\def\Rnm{R_{n-1}}
\def\Rnmm{R_{n-2}}
\def\Rnp{R_{n+1}}
\def\Rnpp{R_{n+2}}
\def\An{A_{n}}
\def\Anm{A_{n-1}}
\def\Anmm{A_{n-2}}

\def\Bnm{B_{n-1}}

\def\Np{N^{-1/3}}
\def\Npp{N^{-2/3}}
\def\dagU{U^\dagger}
\def\mm{\mu^2}
\def\Pnz{P_n(z)}
\def\Pnzi{P_n(1/z)}
\def\pg{\partial_\g}
\def\hn{h_n}
\def\pmu{\partial_\mu}
\def\dx{{d\over dx}}
\def\dxx{{d^2\over dx^2}}
\def\pt{\partial_t}
\def\ptau{\partial_\tau}
\def\expb{e^{-(4/3)x^{3/2}}}
\def\xM{x^M}
\def\xMh{x^{M-1/2}}
\def\eps{\epsilon}

\def\th{\theta}
\def\la{\lambda}
\def\Pint{{\rm P}\int}
\def\Re{\rm Re}
\def\Im{\rm Im}
\def\DFs{\Delta F_s}
\def\tf{\tilde f}
\def\tx{\tilde x}

\REF\GM{D. J. Gross and A. A. Migdal, \PRL {\bf 64} (1990), 717.}
\REF\BrKaz{E.~Br\'ezin and V.~Kazakov, \PL {\bf 236B} (1990), 144.}
\REF\DSh{M. Douglas and S. Shenker, \NP {\bf B335} (1990), 635.}
\REF\doug{M. R. Douglas, \PL {\bf 238B} (1990) 176.}
\REF\gmig{D. J. Gross and A. A. Migdal, \NP {\bf B340} (1990), 333.}
\REF\PSh{V.~Periwal and D.~Shevitz, \PRL {\bf 64} (1990), 1326.}
\REF\PShII{V.~Periwal and D.~Shevitz, \NP {\bf B344} (1990) 731.}
\REF\BDSS{T.~Banks, M.~Douglas, N.~Seiberg and S.~Shenker, \PL {\bf 238B}
(1990), 279.}
\REF\GD{I.~Gel'fand and L.~Dikii, \journal Usp. Matn. Nauk.\  &30 (75) 5.}
\REF\ZabKr{N.~J.~Zabusky, and M.~D.~Kruskal, \PRL {\bf 15} (1965), 240}
\REF\DSS{M.~Douglas, N.~Seiberg and S.~Shenker, \PL {\bf 244B} (1990), 381.}
\REF\Neub{H.~Neuberger, \NP {\bf B340} (1990), 703.}
\REF\GW{D.~J.~Gross and E.~Witten, \PRD {\bf21} (1980), 446.}
\REF\Wadia{S~.Wadia, \PL {\bf 93B} (1980), 403.}
\REF\AbSeg{M.~J.~Ablowitz and H.~Segur, {\it Solitons and the Inverse
Scattering
Transform}, (SIAM, Philadelphia, 1981).}
\REF\MinahanI{J.~A.~Minahan, \PL {\bf 268B} (1991), 29.}
\REF\MyP{R.~C.~Myers and V.~Periwal, ITP preprint, NSF-ITP-91-01, 1991.}
\REF\MinahanII{J.~A.~Minahan, \PL {\bf 265B} (1991), 382.}
\REF\BMP{E.~Br\'ezin, E.~Marinari and G.~Parisi, \PL {\bf 242B} (1990), 35.}
\REF\NumRec{{\it c.f.} W.~Press, B.~Flannery, S.~Teukolsky and W.~Vetterling,
{\it Numerical Recipes,} (Cambridge University Press, Cambridge, 1986).}
\REF\Planar{E.~Br\'ezin, C.~Itzykson, G.~Parisi and J.~B.~Zuber, \CMP {\bf 59}
(1978), 35.}
\REF\Shenker{S.~Shenker, ``The Strength of Nonperturbative Effects in String
Theory", Proceedings of Carg\`ese meeting on Random Surfaces, Quantum Gravity
and Strings, May 1990.}
\REF\GZJ{P.~Ginsparg and J.~Zinn-Justin, ``Action Principle and Large Order
Behavior of Non-Perturbative Gravity'',
Proceedings of Carg\`ese meeting on Random Surfaces, Quantum Gravity
and Strings, May 1990.}
\REF\DavidII{F.~David, \NP {\bf B348} (1991), 507.}

\chapter{Introduction}

The study of matrix models has led to the uncovering of many rich structures.
Not only do they describe two dimensional gravity coupled to conformal
matter\refmark{\GM-\DSh}, at least perturbatively,
but they have also been shown to satisfy a KdV
hierarchy for the Hermitian case\refmark{\doug,\gmig},
and a mKdV hierarchy for the unitary case\refmark{\PSh,\PShII}.

The generic one-matrix model satisfies the differential
equation\refmark{\gmig,\BDSS}
$$x=\sum_{k=0}^\infty (k+1/2)T_kR_k[u],\eqn\Genfloweq$$
where the $R_k[u]$ are the Gel'fand-Dikii potentials\refmark{\GD}.
Changing the coupling $T_k$ induces a flow equation
$${\partial u \over \partial T_k}=R_{k+1}[u],\eqn\genKdV$$
which is the generalized KdV equation.
This equation suggests a possible existance of solitons in the flows
of these theories.

%But in what sense could there be solitons?
Recall that there exist soliton solutions to the KdV equation\refmark{\ZabKr}
$$\pt u(x,t) +u'''(x,t)-6u(x,t)u'(x,t)=0.\eqn\KdV$$
Moreover, this equation also has multiple soliton solutions.
For the multiple solutions, the solitons are able to pass through
one another and are not necessarily of the same size nor moving at the same
velocity.  Likewise, the modified KdV equation
$$\pt u(x,t)+u'''(x,t)-6u^2(x,t)u'(x,t)=0\eqn\mKdV$$
has solutions with these same properties.

But these solitons are disturbances that propagate
over the background, $u(x)=0$.  The flow equations that arise in matrix models
certainly do not have a limit $u(x)=0$ as $T_k\to\pm\infty$.  For instance,
the authors of ref. [\DSS] investigated the flow away from the $m=3$
multi-critical point by turning on the potential $T_2$.  In the limit
$T_2\to-\infty$ they found that the theory flows to the $m=1$ critical point,
while for the other sign, the theory becomes highly oscillatory.  Neither
limit has $u(x)=0$.  It might be possible to see a solitary
wave propagating over a nontrivial background,  but this does not occur for
the case described in [\DSS]. Instead, as $T_k\to\infty$, infinitely many
oscillations develop in a dispersive wave.

In this paper we will describe a matrix-model which has flows with finitely
many
solitary waves.
The model that we will study is a unitary matrix model with a potential tuned
to its lowest multicritical point, which is essentially
QCD on a single plaquette\refmark{\GW,\Wadia}.
To this model we will couple $M$ flavors of nonpropagating bosonic quark terms.
Integrating out the quarks will lead to an effective action with logarithmic
terms.  The effective potential will have some dependence on the rescaled quark
mass term, $\mu$.  We will show that the square root of the specific heat,
$f(x)$, will satisfy a fourth order differential equation, where $x$ is
the rescaled coupling.  This equation is related to the mKdV analog of the
Gel'fand-Dikii equation, which generates the KdV hierarchy.  Moreover,
we will also show that $f(x)$ satisfies a flow equation when $\mu$ is varied.
We will find that the flow equation is not quite the mKdV equation, but
the mKdV equation with a dissipative
term\refmark{\AbSeg} that is not galilean invariant.

By studying these equations numerically, we will see that the flow equation
does indeed describe the propagation of soliton like objects.  We have
found explicit numerical solutions for one and two flavors of quarks,
and the corresponding flow equations describe propagation of one and
two solitary waves respectively.
This will lead us to conjecture that for $M$ flavors of quarks ther eill be $M$
solitary waves in the flow.
Unlike the typical soliton solution that propagates over a background of
$u(x,t)=0$, the waves we find propagate over a background that is a
solution to the Painlev\'e II equation.
The solitary waves die out as $\mu=\infty$ and move off to $x=\infty$
when $\mu=0$; they leave
a nontrivial solution to the Painlev\'e II equation in their wakes.
This will lead us to conclude that the massless quark case has the same
specific
heat as the infinitely massive case.  This remarkable ``duality'' is
nonperturbative in nature, that is to say, it is not evident from the
equations of motion.  Note that from hereon,
we will refer to these solitary waves
as solitons, even though they don't quite meet the Zabusky-Kruskal criteria
for such objects.

We have also studied the limit $M\to\infty$ to see whether or not it
can serve as a reasonable regularized version of pure 2d
gravity\refmark{\MinahanI,\MinahanII}.  We will
show how to calculate the nonperturbative contributions to the specific
heat and will show that they swamp the perturbative behavior found
in the relevant scaling regime.

In section 2 we derive the differential equation for $f(x)$,
as well as its flow equation.  In section 3 we present our numerical
results.  In section 4 we study the  nonperturbative contributions
to the specific heat.  In this section we also show that 2d gravity cannot
be reached from a large $M$ theory.  In section 5 we present our conclusions.

\chapter{Derivation of the Equations}

In this section we derive a differential equation for $f(x)$, the square
root of the specific heat,  where $x$ is the rescaled coupling.  The
derivation uses the method of orthogonal polynomials, using techniques
described by Periwal and Shevitz\refmark{\PSh} and by Neuberger\refmark{\Neub}.

Let us assume that we have a unitary matrix model with $M$ flavors of
bosonic ``quarks'' coupled to the $N\times N$ matrices.  The quark terms
are $N$ dimensional vectors, $\psi_i$ and $\chi_i$,
that transform canonically under
unitary transformations.  Let us choose the potential term to be of the
form
$$N\chi^*_i(\g-U)\psi_i+{\rm c.c.},$$
where $U$ is a unitary matrix.  Since the potential term is quadratic in
$\psi$ and $\chi$, these fields
 can be integrated out to give the effective potential
$$M\tr\log(\g+1/\g -U-U^\dagger),$$
where we have dropped unimportant constant terms.  Notice that potential
becomes unbounded below as $\g\to|1|$, hence we expect critical behavior to
occur for $\g$ near these values.  The complete potential for the lowest
multicritical point is now given as
$$V=Ng\tr(U+U^\dagger)+M\tr\log(\g+1/\g+U+U^\dagger).\eqn\Vpot$$

The partition function is
$$Z=\int D Ue^{-V(U+\dagU)}\eqn\Partfn$$
and is reducible to the diagonal form
$$Z=\oint {dz_i\over 2\pi i z_i}\prod_{i<j}|z_i-z_j|^2e^{-\sum_iV(z_i+1/z_i)}.
\eqn\PartfnII$$
Following [\PSh] and [\Neub],
we define the polynomials
$$P_n(z)=z^n+\sum_{k=0}^{n-1}a_{k,n}z^k\eqn\genII$$
where all $a_{k,n}$ are real and the $P_n$'s satisfy an orthogonality relation
$$\oint d\mu P_n(z)P_m(1/z)=h_n\delta_{n,m},\eqn\orthonorm$$
$$d\mu={dz\over 2\pi i z}\exp(-{Ng}V(z+1/z)).\eqn\dmueq$$
$P_n$ satisfies the recursion relations
$$P_{n+1}(z)=zP_n(z)+R_n z^nP_n(1/z)\eqn\recursion$$
$${1\over h_{n+1}}P_{n+1}(z)={z\over h_n}P_n(z)+
{R_n z^{n+1}\over h_{n+1}}P_{n+1}(1/z)\eqn\recursionII$$
where $a_{0,n}=R_{n-1}$ and $h_{n+1}/h_n=1-R_n^2$.
Let us define $A_n=a_{n-1,n}$ and $B_n=a_{n,n+2}$.  The recursion relations
\recursion\ and \recursionII\ lead to the relations
$$A_n=A_{n-1}+R_nR_{n-1},\eqn\Arec$$
$$B_n=B_{n-1}+R_{n+1}(R_{n-1}+R_nA_{n-1}),\eqn\Brec$$
which will result in simple differential equations when we take the limit
$N\to\infty$.

The next step is to consider the integral
$$\oint d\mu e^{-V(z+1/z)}\partial(z^a P_{n+1}(z)P_n(1/z))$$
for the cases $a=0,1,2$.  Integration by parts leads to the equations
$$(n+1)(h_{n+1}-h_n)=-\oint d\mu V'(z+1/z)(1-1/z^2)P_{n+1}(z)P_n(1/z),
\eqn\intI$$
$$ A_n h_n=-\oint d\mu V'(z+1/z)(1-1/z^2)z P_{n+1}(z)P_n(1/z),\eqn\intII$$
$$ \eqalign{\bigl((n+2)(R_{n+1}R_{n-1}&(1-R_n^2)-A_n^2+2B_{n-1}\biggr)h_n=\cr
&-\oint d\mu V'(z+1/z)(1-1/z^2)z^2 P_{n+1}(z)P_n(1/z).}\eqn\intIII$$

Now in our case, we find that
$$V'(z+1/z)=Ng+{M\over \g+1/\g+z+1/z}.\eqn\genIII$$
It is clear that in order to derive a differential equation from
\intI-\intIII\ we will need to arrange the terms so that we
cancel out the $z+1/z$ term that appears in the denominator.
This is done by multiplying \intII\ by $\g+1/\g$ and adding it to the sum
of \intI\ and \intIII.  After some straightforward but tedious algebra
we are left with the equation
$$\eqalign{2\Bnm&-\An^2+(\gg)\An+(n+2)\bigl(\Rnp\Rnm(1-\Rn^2)-\Rn^2\bigr)
+\Rn^2\cr
&={Ng}\biggl[-\Rn(\Rnp+\Rnm)+\Rnpp\Rnm+\Rnmm\Rnp\cr
&\qquad-\Rnm\Rnp\bigl(\Rnmm\Rnm+\Rnm\Rn+\Rn\Rnp+\Rnp\Rnpp\bigr)\cr
&\qquad+(\gg)(1+\Rnp\Rnm)\biggr](1-\Rn^2)\cr
&\ -M(1+\Rnm\Rnp)(1-\Rn^2).}\eqn\bigrel$$

We now take the limit $N\to\infty$ and set $g=1/2$. Consider the
rescaled variables
$$\eqalign{x&=(N-n-M)\Np\cr \Rn&=f(x)\Np}\qquad\qquad
\eqalign{\g&=-1-\mu\Np\cr \An&=N/2+g(x)\Np}$$
$$2\Bnm=\An^2+2\An-\Rn^2-N-M+2h(x)N^{-1}.$$
After substituting these variables into \bigrel, we are led to the equation
$$\eqalign{2h-\mu^2 g=x\bigl(ff''&-(f')^2-f^4\bigr)
+\half ff''''-f'f'''+\half(f'')^2\cr
&-5f^3f''+2f^6 -{\mu^2\over2}\bigl(ff''-(f')^2-f^4\bigr).}\eqn\sceqI$$
The rescaled versions of \Arec\ and \Brec\ are respectively
$$g'=-f^2,\qquad\qquad{\rm and}\eqn\sceqII$$
$$h'={3\over2}f^4-ff''+\half(f')^2.\eqn\sceqIII$$
The last equation is found by observing that
$$\eqalign{\An^2-\Anm^2&=(\An+\Anm)\Rn\Rnm\cr
&=2\Anmm\Rn\Rnm+(\Rn\Rnm)^2+2\Rn\Rnm\Rnm\Rnmm.}$$

Taking derivatives on both sides of \sceqI\ and making the substitutions
given by \sceqII\ and \sceqIII, one finds the equation
$$\biggl({H''-\mu^2H\over f}\biggr)'=4fH',\eqn\lineq$$
where $H$ is defined by
$$H=f''-2f^3+2xf.\eqn\Heq$$
$H$ is the Painlev\'eII operator, and it is relatively straightforward
to show that
for potentials leading to higher multicritical behavior,
the generic form of $H$ is
$$H=\sum_i T_i {\cal R}_i +2xf$$
where the $T_i$ are essentially couplings and the ${\cal R}_i$ are
the unitary analogs of the Gel'fand-Dikii potentials.  Moreover,
\lineq\ is the generator of these potentials.

$M$ does not explicitly appear in \lineq, but rather arises as
an integration constant.  Multiplying both sides of \lineq\ by
$(H''-\mu^2H)/f$ and then integrating gives the equation
$$(H''-\mm H)^2=4f^2((H')^2-\mm H^2 +C).\eqn\maineq$$
Notice that when $\mu$ is zero, \maineq\ has the solution $H=C$, the generic
Painlev\'e II equation.
The integration constant $C$ can be determined by looking at the perturbative
solutions to \maineq.  In particular, the asymptotic expansion derived
from \maineq\ for large $x$ is
$$f(x)=x^{1/2}-{C^{1/2}\over 4\mu x}+{C^{1/2}\mm\over 16\mu x^2}+
{\rm O}(1/x^{5/2}),$$
but it is known that for $\mu=0$, the leading order expansion
is\refmark{\MinahanI-\MinahanII}
$$f(x)=x^{1/2}-{M\over 2x}+{\rm O}(x^{-5/2}).\eqn\asymp$$
Hence we see that $C=4\mm M^2$.

We conclude this section by deriving a flow equation for $f(x)$ as a
function of $\mu$.  The flows are basically determined by ward identities
of the matrix model partition function.
We can take a derivative with respect to $\g$ on both sides of \orthonorm,
giving the equation
$$-{M\over\g}\oint d\mu {2\g+z+1/z\over \g+1/\g+z+1/z}\Pnz\Pnzi=\pg h_n.
\eqn\ward$$
We can also do the same thing after inserting $z$ and $1/z$ into \orthonorm,
giving the equation
$$\eqalign{-{M\over\g}\oint d\mu &{2\g+z+1/z\over \g+1/\g+z+1/z}z\Pnz\Pnzi\cr
&\qquad+h_n\pg\An- h_n\Rn\pg\Rnm=-\pg (h_n\Rn\Rnm),}\eqn\wardII$$
and an analogous equation for the $1/z$ insertion.
Combinining all three equations, we can deduce the expression
$$-{2M\over\g}(\g-\Rn\Rnm)=(\g+1/\g-2\Rn\Rnm){\pg\hn\over\hn}-2\Rnm\pg\Rn
-2\pg\An.\eqn\comboward$$
Since $\hn$ is given by
$$\hn=h_0\prod_{i=0}^{n-1}(1-R_i^2),$$
we can derive the recursion relation for its derivative
$${1\over h_{n+1}}\pg h_{n+1}-{1\over \hn}\pg\hn={-2\Rn\pg\Rn\over 1-\Rn^2}
\eqn\phrec$$
Taking a difference of \comboward\ and invoking \Arec\ and \phrec\ leads to
the equation
$$\eqalign{{2M\over\g}(\Rnp-\Rnm)=&(\g+1/\g-2\Rnp\Rn)
\ {-2\Rn\pg\Rn\over1-\Rn^2}\cr
&-2\Rn(\Rnp-\Rnm){\pg\hn\over\hn}-2\Rn\pg(\Rnp+\Rnm).}\eqn\diffeqI$$
We can get rid of this last $\pg\hn/\hn$ term by dividing \diffeqI\ by
$\Rn(\Rnp-\Rnm)$, taking another difference and then multiplying by
$\Rnp-\Rnm$.  Hence we find
$$\eqalign{0=&{2\Rn(\Rnp-\Rnm)\pg\Rn\over 1-\Rn^2}
-{\Rnp-\Rnm\over\Rnpp-\Rn}\cr
&\qquad\times\biggl(
(\g+1/\g-2\Rnpp\Rnp){\pg\Rnp\over1-\Rnp^2}+\pg(\Rnpp+\Rn)
\biggr)\cr
&\qquad+(\g+1/\g-2\Rnp\Rn){\pg\Rn\over 1-\Rn^2}+\pg(\Rnp+\Rnm).}\eqn\diffeqII$$
Taking the double scaling limit then leads to the equation
$$\eqalign{0=4ff'\pmu f+\mm\pmu f'+8ff'\pmu f +&4f^2\pmu f'-\pmu f'''\cr
&-{f''\over f'}(\mm\pmu f +4f^2\pmu f -\pmu f''),}\eqn\floweq$$
which, after dividing through by $f'$ and changing the order of derivatives,
gives
$$\biggl(4f+\dx{1\over f'}(\mm+4f^2-\dxx)\biggr)\pmu f=0\eqn\flowII$$
Hence the flow equation is defined by a linear operator and the flows
are determined by finding the eigenfunctions with zero eigenvalue.

Now let $\pmu f=\psi$.  If we substitute $\int\psi$ for $H$ in \lineq, then it
is straightforward to show that the equation reduces to the flow equation,
\flowII.
Therefore, we find that
$$\pmu f=C H',\eqn\floweqIII$$
where $H$ is given by \Heq.  The constant $C$ can be determined by looking at
the asymptotic expansion for $f$.  The leading behavior is easily determined
from \maineq, giving
$$f=\sqrt{x}-{M\over2x}+{M\mm\over16x^2}+...,$$
and hence the leading term for $\pmu f$ is $2M\mu/16x^2$.
The leading term for $H'$ is $M\mm/4x^2$, thus $C=1/2\mu$.  Letting
$t=\log\mu$,
we can then write the flow equation as
$$\pt f=\half(f''-2f^3+2xf)'.\eqn\flow$$

Equation \flow\ is almost the modified KdV equation, the difference being that
there is extra term, $xf'+f$.  This term does two things--- it breaks
galilean invariance and it dissipates wave motion.
Nevertheless,  this equation has soliton like solutions, albeit
solitons that die out.  Our numerical results will exhibit this
behavior.

\chapter{Numerical Results}

A relaxation method was used to find solutions to \maineq\ with the asymptotic
behavior given by \asymp.
Following the work of [\BMP,\DSS] we consider the diffusion equation
$$\ptau f=D(f),\eqn\diffusion$$
where $D(f)$ is
$$D(f)=(H''-\mm H)^2-4f^2((H')^2-\mm(H^2-4M^2)).$$
If $f$ relaxes to a $\tau$ independent solution, then we have found a solution
to \maineq.  One starts with a test function with the desired
asymptotic behavior and lets it evolve.
This is done by using the recursion
$$f_{\rm new}=f_{\rm old}+hD(f_{\rm old}).$$
However, $h$ cannot be chosen too large, otherwise the system destabilizes.

The problem with this approach is that a solution is not
always reached,  sometimes $f$ blows up.  In fact for $\mm$ less than a
certain value it turned out to be nearly impossible to find stable solutions.
However if we can find a solution for a given value of $\mm$, we can then
employ the flow equation \flow\ to find approximate solutions for smaller
$\mm$.

A starting function was generated by considering solutions to
the Painlev\'e II equation
$$g''=2g^3-2xg+2M.\eqn\PII$$
It turns out that for $M<1/2$ there is a solution to \PII\ with
asymptotic behavior $g\approx+\sqrt{x}$ as $x\to\infty$,
which is free of poles.
The asymptotic behavior of this solution as $x\to-\infty$ is $g\approx M/x$.
$M$ was set to $M=0.49$ and a
numerical solution was found
by using the relaxation method, where the starting function was chosen to be
a solution
to the algebraic limit of \PII.  The algebraic solution with the desired
asymptotic behavior has a jump at the critical point $x=3(M/2)^{2/3}$,
but this did not prove to be a hindrance in finding a solution to \PII.

The resulting solution of \PII\ was then used as a starting function for
\diffusion.
Since the asymptotic behavior of a solution to \PII\ is slightly different than
that for \maineq, the endpoints of $f(x)$ were allowed to move for a large
number of iterations, but were eventually fixed at $x=\pm20.0$.
The difference in asymptotic behavior is most apparent
in the strong coupling regime ($x\to-\infty$) where the behavior for nonzero
$\mu$ is $f\sim e^{\mu x}$.
With $M$ still set to $0.49$, we were able to reach a solution for $\mu=1$.
This solution was then used as the starting function for different values
of $M$ and $\mu$.

\FIG\FigI{Flows of $f(x)$ for $M=1$ and different values of $\mu$.  $\mu$
ranges
from 2.1 down to $1.1\times 10^{-3}$.
Also shown is a plot of $f(x)$ for $M=0$.}
Figure 1 shows the results for $M=1$.  The solution for $\mu=2.1$ was found
using relaxation  and the lower values of $\mu$ were reached using the
flow equation.  The flow was calculated using a fourth
order Runge-Kutta routine.
Moreover, after each step in the flow,
a conjugate gradient routine\refmark{\NumRec} was
used to minimize the value of $\sum_i(D(f(x_i)))^2$.
We used two different lattice
spacings ($\Delta x=.3$ and $.1$) and two different flow steps ($\Delta t=-.01$
and $-.001$) with no qualitative difference in results.
The flow is glacial because of the log dependence of $\mu$;
the values of $\mu$ plotted
here start at $\mu=2.1$ and end at $\mu=1.1\times 10^{-3}$
Figure 1 also has a plot of
a solution to \PII\ with $M=0$ and weak coupling behavior $- \sqrt{x}$.
Since the specific heat of the theory is given by $-f^2(x)$,
it is clear that as $\mu$ approaches zero, the behavior of $f(x)$ in the
strong coupling regime approaches that of pure QCD.
The only region where $f(x)$ differs markedly from this behavior is
in the region where $f(x)$ increases sharply.  Let us contrast this
with the behavior of $f(x)$ as $\mu\to\infty$.  In this case the quark
terms should decouple and the theory should also behave like pure QCD.
In this case $f(x)$ approaches the solution of \PII\ with $M=0$ but with
asymptotic behavior $\sqrt{x}$.  Hence we see that specific heat in the
scaling regime is identical for the zero-mass and the infinite mass case.

Let us next compare the specific heat for the case $M=1$ and assorted values
of $\mu$ with the specific heat of pure large $N$ QCD.
\FIG\FigII{Plots of $f^2_{\rm QCD}(x)-f^2_\mu(x)$, with $M=1$ and $\mu$
ranging from 2.1 (bottom) to $1.1\times 10^{-3}$(top).}
Figure 2 shows
the difference in specific heats, $f^2_{\rm QCD}(x)-f^2_\mu(x)$, for values of
$\mu$ ranging from 2.1 down to $1.1\times10^{-3}$.
The figure suggests a soliton propagating in the negative $x$ direction as
the value of $\mu$ increases, but with dissipating size.
When $\mu=0$, the soliton is off at positive infinity  with no wake
in the finite $x$ regime.
As the wave travels in the $-x$ direction it eventually dies out, leaving
pure QCD behavior for the specific heat.

This dual structure between the two mass extremes is quite surprising, at least
from the standpoint of the original matrix model partition function.  It is
a nonperturbative effect and is not seen in a saddle point calculation of the
eigenvalues.  The peak that appears in the specific heat is similar
to the Schottky anomaly, which is the result of tunneling in a two
well system.  If the first $N-1$ eigenvalues relax to the ground state, then
the $N^{\rm th}$ eigenvalue will have an effective potential with two wells,
provided that $\mu$ is small enough.  The presence of this second well will
result in a relatively large density of states around some corresponding
energy,
leading to a peak in the specific heat.  The effects of these wells will be
discussed further in the next section.

We have also succeeded in finding a solution for $M=2$ and $\mu=3.5$,
although only for lattice sizes larger than or equal to $\Delta x=0.3$.
With this solution we then flowed
to lower values of $\mu$,  giving the results shown in figure 3.
\FIG\FigIII{Flows of $f(x)$ for $M=2$ and different values of $\mu$.
$\mu$ ranges from 3.5 down to $10^{-2}$.}
Here we see
that an extra bump develops as $\mu$ gets small.  Unfortunately,
we were unable to reach values of $\mu$ below $.01$ without developing
large errors in the value of $\sum_i (D(f_i))^2$.   However, this seemed
far enough to clearly see the second peak.
\FIG\FigIV{Plots of $f^2_{\rm QCD}(x)-f^2_\mu(x)$, with $M=2$ and $\mu$
ranging from 3.5 (bottom) to $1.1\times 10^{-1}$(top).}
The peaks are quite clear in
figure 4, where we have graphed $f^2_{\rm QCD}-f^2_\mu$ for values of $\mu$
between 3.5 and 0.1.  In this case there appear to be two soliton peaks
where one clearly propagates in the $-x$ direction and eventually dies out.
The second peak definitely moves slower than the first, although it is not
clear from our data if it even moves at all.  However, we deem it likely that
this peak eventually moves off to infinity as $\mu$ approaches zero.

It is natural to postulate that a system with $M$ bosonic flavors will have $M$
of these peaks in its specific heat.  It is reasonable to expect that the
peaks are due to
classical configurations corresponding to some sort of instanton processes,
which are clearly a result of the quark terms in the effective action.
A question
worth asking is whether these configurations are also present in two
dimensional
lattice QCD with propagating quarks.  Recall that the theory we have been
considering is QCD
restricted to one plaquette.  If there are no quarks in the theory, then this
is
equivalent to QCD on a lattice, since in the full lattice theory
the plaqettes decouple.  But once propagating quarks are
introduced the plaquettes become coupled.
However it is possible that the peak structure found
here will have some remnant in the full two dimensional theory,
if, for example, the two dimensional theory has soliton solutions that are
localized in space.  This problem is left for future work.

\chapter{Gravity is Unreachable}

In [\MinahanI] it was argued that as the number of flavors $M$ becomes large,
then under a rescaling, the Painlev\'e II equation reduces to the Painlev\'e I
equation.  This suggests that pure two dimensional gravity can be reached
from two dimensional QCD.  However, in [\MinahanII] it was shown that the
theory with the correct asymptotic behavior will have poles in $f(x)$
if we insist that $f(x)$ is real.  This was shown to contradict
the Schwinger-Dyson equations.  It was then suggested that under a suitable
regularization we could remove the poles and reach pure gravity.
The poles were arising because the effective matrix potential was unbounded
below.  This is a consequence of setting $\mu$ to zero.  If we were to
bound this by introducing a nonzero $\mu$, then we can consider the
simultaneous
limit $\mu\to0$ and $M\to\infty$ and check whether gravity is reached.  We
will now show that this won't occur.

When $M$ is large we can set $\eps=2/M$ and
define $\tx$ and $\tf(\tx)$ such that
$$\eqalign{x&=3\eps^{-2/3}+(3/4)^{1/5}\eps^{2/15}\tx\cr
f&=\eps^{-1/3}+(1/18)^{1/5}\eps^{1/15}\tf.}\eqn\rescale$$
Then to leading order in $\eps$ the Painlev\'e II equation reduces to
Painlev\'e I.  Therefore, in order that the regularized theory
reach pure gravity,
the solutions that satisfy the boundary conditions must pass sufficiently close
to the critical point $x=3\eps^{-2/3}$ and $f=\eps^{-1/3}$.

Let us next consider the nonperturbative behavior that can be derived from
\maineq.  Define $f(x)$ as $f_{\rm asy}(x)+g(x)$, where $f_{\rm asy}$ is
the asymptotic expansion in $1/x$ and $g(x)$ is a subdominant term. To leading
order in its asymptotic expansion, $H(x)$ is given by
$$\eqalign{H(x)&=2M-M\mm/4x+...+g''(x)-4xg(x)+...\cr
&=2M-M\mm/4x+...+h(x)}\eqn\Hasymp$$
Plugging this into \maineq\ leads to a linear equation, which to first
approximation is
$$-4\mm Mh''= 4(x-M/x^{1/2})((M\mm/2x)h'-4M\mm h).\eqn\hlinear$$
We have assumed that $\mu$ is small and thus higher order
terms in $\mu$ can be dropped.  Letting $h=\rho(x)\expb$, we can use a
WKB approximation to derive the equation
$$4\rho'+\rho=-\rho+4M\rho,\eqn\WKBeq$$
thus giving $\rho=C \xMh$, where $C$ is a constant in $x$ that
depends on $M$ and $\mu$.  Now $g(x)$ satisfies the approximate equation
$$h(x)=g''(x)-(4x-(6M/x^{1/2}))g(x).\eqn\hglinear$$
This leads to  $g(x)=r(x)\expb$ where $r(x)$ approximately satisfies the
equation
$$\rho(x)=-4x^{1/2}r'(x)-r(x)/x^{1/2}+6Mr(x)/x^{1/2}.\eqn\rhoeq$$
Hence $r(x)$ can be approximated as $r(x)\approx C\xM/(2M-1)$

We can derive the $\mu$ dependence of $C$ using the flow equation.  From
\flow\ we find that
$$\pt g=\half h'\approx-(2M-1)g,$$
from which we derive that $C\sim e^{-(2M-1)t}=1/\mu^{2M-1}$, and
$$g(x)\approx{\kappa\mu(x/\mm)^M\over 2M-1}\expb.\eqn\geq$$
$\kappa$ depends only on $M$.
{}From this form of $g(x)$ we see that when $\mu\to0$, the perturbative
expansion will be wiped out by nonperturbative effects, if $M\ge1/2$.
However, for finite $M$, if $\mu$ is small then there is a large range
of values for $x$ where the perturbative expansion of the Painlev\'e II
equation is accurate to large order in $1/x$.
This is because the nonperturbative piece only depends on a power of $\mu$.
Hence, while nonpertubative
effects ruin the limit to Painlev\'e II, the effect is quite gentle.

If we now consider the large $M$ limit, it is possible that these
nonperturbative effects will be gentle enough that they won't
destroy the Painlev\'e I
behavior, perhaps because the overall coefficient in front of the perturbative
piece is small.  That is to say, we can choose $\mu$ to be small and yet have
$f(x)$ pass sufficiently close to the critical point.
We now demonstrate that nonperturbative effects do indeed destroy the limit.

In order to understand this, let us consider the origin of the nonperturbative
contribution for the lowest multicritical point of the unitary matrix models.
The square root of the specific heat is a solution to the Painlev\'e II
equation
with $M=0$.  The perturbative behavior can be determined from the distribution
of eigenvalues on a circle.  The potential is given by
$$V(\th)=2Ng\sum_i\cos\th_i-\sum_{i<j}\log(\sin^2\half(\th_i-\th_j)),\eqn\PotI$$
and the equations of motion are
$$0=-2gN\sin\th_i-N\sum_{j\ne i}\cot\half(\th_i-\th_j).\eqn\PoteqI$$
Defining $x=i/N$, $dx=1/N$ gives the equation
$$0=-2g\sin\th(x)-\Pint_0^1dx \cot\half(\th(x)-\th(x')),\eqn\PoteqII$$
where the P in front of the integral indicates the principle value.
Introducing the density of eigenvalues $u(\la)=dx/d\la$, leads to the equation
$$0=-2g\sin\th-\int_{-a}^{a}d\la\cot\half(\th-\la),\eqn\PoteqIII$$
where $\pm a$ are the end points of the eigenvalue distribution.

The density of states can be determined in a way analogous to the method used
in [\Planar].  We define the function $F(\th)$,
$$F(\th)=\int_{-a}^{a}d\la\cot\half(\th-\la),\eqn\Feq$$
which is analytic everywhere on the strip $-\pi<\Re\th<\pi$, except for
a cut on the real line between $-a$ and $a$.  Using the equations of motion
and the fact that $F(\th)\to\mp i$ as $\Im\th\to\pm\infty$, results in
$$u(\la)={g\over\pi}\sqrt{2(\cos\la-\cos a)}\cos(\la/2).\eqn\density$$
$a$ is related to $g$ by
$$g(1-\cos a)=1.\eqn\grel$$
This density of eigenvalues then gives the perturbative expansion to the
theory.

However, it is clear that the potential has a saddle point if
one of the eigenvalues is sitting at $\th=\pi$. Hence we look for a solution
where an eigenvalue is fixed at this position and the others are allowed
to seek a local minimum.  The potential for this configuration is given
by
$$\eqalign{V(\th)=&2gN\cos\th-\sum_{i\ne1}\log(\sin^2\half(\th-\th_i))\cr
&+2Ng\sum_{i\ne1}\cos\th_i-\sum_{1<i<j}\log(\sin^2\half(\th_i-\th_j)),}
\eqn\Potlm$$
where $\th=\pi$.  But this is the potential for a $U(N-1)$
system with one fermion flavor  and we know what its perturbative
expansion looks like.  Hence we can determine the nonperturbative contribution
by comparing the perturbative behavior of these two theories.

The partition function for a unitary matrix model is given by
$$Z_N=e^{-F_N}=N!\prod_{i=1}^N(1-R_i^2)^{N+1-i}.\eqn\ZNpart$$
If we assume that $R_i^2=i/N$, which is the tree level contribution,
then $F_N$ is given by
$$\eqalign{F_N&=-\log(N!)-\sum_i(N-i)\log(1-i/N)\cr
&\approx -\log(N!)-{(N+1)^2\over2gN}
\bigl(\log({N+1\over2gN})-3/4\bigr)+C.}\eqn\FreeEn$$
Comparing this with $F_{N-1}$, we find that these two free energies
differ from each other by $\log(N)-1$ if $g$ is in the double scaling region.
But we also need to take into account the difference in the free energy
from the scaling region.  This contribution satisfies
$F''_s=f^2(x)$.  For the zero quark case, the perturbative expansion
of $f(x)$ is
$$f(x)=x^{1/2}-{1/16\over x^{5/2}}+...,\eqn\zqasymp$$
while for the one fermion case, $f(x)$ is
$$f(x)=x^{1/2}+{M/2\over x}-{3M^2/8+1/16\over x^{5/2}}+...,\eqn\fasymp$$
with $M=1$.
These expansions are each easily found from the Painlev\'e II equation.
The difference in $f^2(x)$ for the two situations is then
$$\delta f^2(x)={1\over x^{1/2}}-{1/4\over x^2}+...\eqn\delspheat$$
and therefore the difference in the free energy is
$$\Delta F_s=(4/3)x^{3/2}+\half\log x\Npp+...\eqn\delFreeEn$$
The appearance of $\Npp$ is the usual situation of unscaled variables
appearing in the logarithm terms.

Given the above, we see that the nonperturbative contribution to the partition
function is the perturbative contribution multiplied by the factor
$$C{1\over x^{1/2}\Np}\expb \int d\th e^{-\th^2 V''(\pi)/2}.\eqn\npfact$$
where $C$ is a constant of order unity.
The $\log N$ term that appears in the difference of the two free energies
disappears because any one of the $N$ eigenvalues can sit
at the top of the potential.  Strictly speaking, the density of eigenvalues
used in evaluating $V''$ should be that for the $M=-1$ system.  For a nonzero
value of $M$, $u(\la)$ is modified to
$$\eqalign{u(\la)&={1\over\pi}\biggl(g-{(M/N)(\g^{1/2}+1/\g^{1/2})\over
(\g+1/\g+2\cos\la)\sqrt{\g+1/g+2\cos a}}\biggr)\cr
&\qquad\qquad\times\sqrt{2(\cos\la-\cos a)}\cos(\la/2).}\eqn\Mdensity$$
However, the contribution coming from the term multiplying $M$
will be suppressed by a factor of $1/x^{3/2}$, so we can ignore it.
Taking two $\th$ derivatives of \PoteqIII\ gives
$$V''(\pi)=-2gN+{gN\over2\pi}\int_{-a}^a d\la {\sqrt{2(\cos\la-\cos a)}\over
\cos (\la/2)}.\eqn\Vppeq$$
The integral that appears in \Vppeq\ can be evaluated, satisfying
$$\int_{-a}^a d\la {\sqrt{2(\cos\la-\cos a)}\over \cos (\la/2)}
=\pi(1-\cos(a/2)), \eqn\intrel$$
and hence leading to $V''(\pi)=2gN\cos (a/2)$. Letting
$g=(1+x\Npp)/2$ and using \grel, we find that $V''(\pi)=-x^{1/2}N^{2/3}$.
Since $V''$ is negative, the integral in \npfact\ is not bounded below.
But in any case, the limits of integration should be bounded such that
the argument of the exponent is not much greater than unity, since the
eigenvalue will eventually meet a repulsion from the other eigenvalues.
Hence the nonperturbative contribution to the free energy is
$$F_{np}=C x^{-3/4}\expb,\eqn\Freenp$$
and thus its contribution to $f(x)$ to leading order is
$$f_{np}=C x^{-1/4}\expb.\eqn\specheatnp$$
This is precisely the behavior that one would find for a WKB solution
of \PII.

As an aside, we note that for the hermitian case, if there is a saddle point
such that one eigenvalue is sitting at the critical point,
then the same analysis could be applied to determine
the nonperturbative behavior.  In particular, this would imply that the
free energy of the disk plays some role in this behavior.
For the unitary case, the dominant contribution to the nonperturbative
piece comes from the free energy of the disk, which is proportional to
$1/g$, where $g$ is the renormalized coupling.
But actually, the hermitian matrix models either have no saddle points,
which is the case for the odd potentials, or they are unbounded below.
However, it is true that for the hermitian case, the strength of the
nonperturbative piece\refmark{\Shenker,\GZJ} is proportional to $e^{-C/g}$.
Moreover, it was shown by David\refmark{\DavidII} that the
nonperturbative piece for the pure gravity case is a
consequence of an eigenvalue sitting at the top of a potential.
Hence it is possible that the free energy of the disk will play some role
in determining the nonperturbative behavior of a generic string theory.
We won't have anything more to say about this subject.
%%%  The next paragraph is not really correct
%One could speculate whether this phenomenon
%might persist in more complicated string theories.
%Shenker has shown\refmark{\Shenker} that for generic string theories, the
%strength of nonperturbative effects does behave as $e^{-C/g}$, so at least
%our speculation is consistent with this observation.
%But, if this speculation were true, we would learn that, say,
%for the critical bosonic string theory, these nonperturbative effects
%are completely suppressed, since the free energy from the disk is infinite.

Now let us turn to the situation with $M$ flavors of bosons
and let us assume the $\mu$ is small compared to $1/x$.
If $\mu$ is small enough then the saddle point will actually be a local
minimum.
The effective potential for one eigenvalue fixed at $\th=\pi$ is
$$\eqalign{V(\th)=&2gN\cos\th+\log(\g+1/\g+2\cos\th)
+(M-1)\sum_{i\ne1}\log(\sin^2\half(\th-\th_i))\cr
&\qquad+2Ng\sum_{i\ne1}\cos\th_i-\sum_{1<i<j}\log(\sin^2\half(\th_i-\th_j)),}
\eqn\Potlmm$$
which is the potential for a theory with $M-1$ flavors, if $\th$ is fixed at
$\pi$.
The difference in free energies between these two theories is again found
from the differences in $f^2(x)$, which is
$$\eqalign{\Delta f^2(x)&=(M-(M-1))/x^{1/2}+(M^2-(M-1)^2)/2x^2+...\cr
&=1/x^{1/2}+(M-1/2)/x^2+....}\eqn\delspecheatII$$
{}From this we find the difference in free energies to be
$$\Delta F_s=(4/3)x^{3/2}-(M-1/2)\log (x\Npp)+... .\eqn\DelFreeEnII$$
We should also add the term $M\log(\mm\Npp)$ to $\Delta F_s$, which
is the contribution to the potential from the eigenvalue sitting at $\th=\pi$.
The main contribution to $V''(\pi)$ comes from the second term on the r.h.s.
of \Potlmm, thus $V''(\pi)\approx2/(\mm\Npp)$.  Unlike the previous example,
$V''$ is positive and thus this eigenvalue sits in a well. Hence we should
expect peaks to appear in the specific heat.
Putting this all together, we find that the nonperturbative contribution to
the free energy is
$$F_{np}=C{x^{M-1/2}\over M^{1/2}\mu^{2M+1}}\expb\eqn\FreenpII$$
and therefore the contribution to $f(x)$ is
$$f_{np}=C{x^M\over M^{1/2}\mu^{2M+1}}\expb.\eqn\specheatnpII$$
Hence the behavior matches the subdominant contribution to the asymptotic
expansion of \maineq.

It is clear from these examples that in order for the nonperturbative behavior
to avoid swamping the perturbative expansion,
it is necessary that $\DFs$ diverges
when the scaled variable, in this case $x$, becomes large.  At the same time
we still want to remain in the scaling regime.  To this end, consider what
happens in the large $M$ limit, with the rescaling of variables
given by \rescale.
Given an $x$ there is a corresponding $\tx$, depending on the value of $M$.
If we fix $x$, but decrease $M$ by 1, then $\tx$ shifts by
$\tx\to\tx+2(4\eps/3)^{1/5}$.  \ie\ as $M\to\infty$, the shift in
$\tx$ approaches zero.  Now in order
for the limit $M\to\infty$  to be able to approach 2d gravity, it needs
to have a sensible perturbative expansion in $\tx$.
However, if we assume that shifting $M$ by 1 cannot affect this perturbative
expansion, and since the shift only changes $\tx$ by an infinitesimal
amount, then it must be the case
that a nonperturbative contribution coming from an eigenvalue sitting at
$\th=\pi$ must contribute a piece of order unity if $\mu\sim1$.
In other words, the nonperturbative piece is not suppressed by a factor of
$e^{1/g}$, where in this case, $g=\tx^{-5/4}$.
Hence it must be true that as $\mu\to0$, the curve $f(x)$ must be far from the
critical point  and therefore 2d gravity won't be reached
using this regularization.

\chapter{Discussion}

As we have already mentioned at the end of section 3, an important
question is to what extent the results presented here extend to
the full two dimensional lattice QCD.  For instance, we might wonder
whether these solitons remain in this extended theory.  If they correspond
to localized objects, then it would seem likely that there is a remnant in
the full 2d theory, because their structure appears at the single plaquette
level.
If there is a remnant in the two dimensional theory, then it might turn out
that there is an interesting relation between the solitons one finds in the
flows and the solitons that arise as nontrivial solutions of
the classical field equations.

There is also a question of integrability.  We have been rather cavalier
in referring to these solitary waves as solitons.  In its strict sense,
the name soliton, coined by Zabusky and Kruskal, implies that two such
objects will pass through one another, such that out at infinity, the
only signature that they leave on one another is a phase shift.  This
is a direct consequence of the integrability of the KdV equation.  For
the waves described here they die out while they are passing through
each other, so it is hard to see if there are infinitely many conserved
charges in the theory.  However, it is possible that there is a notion
of integrability in this system and an inverse scattering transform can
be found to solve the problem of wave interaction.
Perhaps there is a change of variables where such a structure would be
made more transparent.
In any event, the fact that this flow equation shows up
in a solvable model lends credence to this supposition.

\ack{It is a pleasure to thank S. Ullah for many helpful discussions about
numerical algorithms.}

\refout
\endpage
\figout

\endpage
\end